# Experimental discovery of a topological Weyl semimetal state in TaP

Photoemission established tantalum phosphide as a Weyl semimetal, which hosts exotic Weyl fermion quasiparticles and Fermi arcs.

Su-Yang Xu*†,[1] Ilya Belopolski*,[1] Daniel S. Sanchez*,[1] Chenglong Zhang,[2] Guoqing Chang,[3,4] Cheng Guo,[2] Guang Bian,[1] Zhujun Yuan,[2] Hong Lu,[2] Tay-Rong Chang,[5] Pavel P. Shibayev,[1] Mykhailo L. Prokopovych,[6] Nasser Alidoust,[1] Hao Zheng,[1] Chi-Cheng Lee,[3,4] Shin-Ming Huang,[3,4] Raman Sankar,[7,8] Fangcheng Chou,[7] Chuang-Han Hsu,[3,4] Horng-Tay Jeng,[5,8] Arun Bansil,[9] Titus Neupert,[10] Vladimir N. Strocov,[6] Hsin Lin,[3,4] Shuang Jia,[2,11] and M. Zahid Hasan†[1,12]

1. Laboratory for Topological Quantum Matter and Spectroscopy (B7), Department of Physics, Princeton University, Princeton, New Jersey 08544, USA
2. International Center for Quantum Materials, School of Physics, Peking University, China
3. Centre for Advanced 2D Materials and Graphene Research Centre National University of Singapore, 6 Science Drive 2, Singapore 117546
4. Department of Physics, National University of Singapore, 2 Science Drive 3, Singapore 117542
5. Department of Physics, National Tsing Hua University, Hsinchu 30013, Taiwan
6. Paul Scherrer Institute, Swiss Light Source, CH-5232 Villigen PSI, Switzerland
7. Center for Condensed Matter Sciences, National Taiwan University, Taipei 10617, Taiwan
8. Institute of Physics, Academia Sinica, Nankang, Taipei R.O.C. Taiwan 11529
9. Department of Physics, Northeastern University, Boston, Massachusetts 02115, USA
10. Princeton Center for Theoretical Science, Princeton University, Princeton, New Jersey 08544, USA
11. Collaborative Innovation Center of Quantum Matter, Beijing,100871, China
12. Princeton Center for Complex Materials, Princeton Institute for the Science and Technology of Materials, Princeton University, Princeton, New Jersey 08544, USA

* These authors contributed equally to this work.
† suyangxu@princeton.edu and mzhasan@princeton.edu

# ABSTRACT

Weyl semimetals are expected to open up new horizons in physics and materials science because they provide the first realization of Weyl fermions and exhibit protected Fermi arc surface states. However, they had been found to be extremely rare in nature. Recently, a family of compounds, consisting of TaAs, TaP, NbAs and NbP was predicted as Weyl semimetal candidates. Here, we experimentally realize a Weyl semimetal state in TaP. Using photoemission spectroscopy, we directly observe the Weyl fermion cones and nodes in the bulk and the Fermi arcs on the surface. Moreover, we find that the surface states show an unexpectedly rich structure, including both topological Fermi arcs and several topologically-trivial closed contours in the vicinity of the Weyl points, which provides a promising platform to study the interplay between topological and trivial surface states on a Weyl semimetal's surface. We directly demonstrate the bulk-boundary correspondence and hence establish the topologically nontrivial nature of the Weyl semimetal state in TaP, by resolving the net number of chiral edge modes on a closed path that encloses the Weyl node. This also provides, for the first time, an experimentally practical approach to demonstrating a bulk Weyl fermion from a surface state dispersion measured in photoemission.

**INTRODUCTION**

Recent developments in low-energy condensed matter physics have established a new route to study fundamental aspects of high-energy physics (*1, 2*). This can in turn help scientists to design better materials leading to new technologies and devices. The experimental realization of Dirac fermions in crystalline solids such as graphene and topological insulators has become one of the main themes of condensed matter physics in the past ten years (*3-6*). A particle that is relevant for both high energy and condensed matter physics is the Weyl fermion (*7*). Weyl fermions may be thought of as the building blocks for any fermion. Therefore, one-half of a charged Dirac fermion of a definite chirality is called a Weyl fermion. Although Weyl fermions play a crucial role in quantum field theory, they are not known to exist as fundamental particles in vacuum.

Recent advances in topological condensed matter physics provided new perspectives into realizing this elusive particle (*8-11*). It was proposed that an exotic type of topologically non-trivial metal, the Weyl semimetal, can host Weyl fermions as its low energy quasiparticle excitations. The resulting singly degenerate bands contain relativistic and massless states that touch at points, the Weyl nodes, and disperse linearly in all three momentum space directions away from each Weyl node. Weyl fermions have distinct chiralities, either left-handed or right-handed. In a Weyl semimetal crystal, the chiralities of the Weyl nodes gives rise to topological charges, which can be understood as monopoles and anti-monopoles of Berry flux in momentum space. The separation of the opposite topological charges in momentum space directly leads to the topologically nontrivial phase in Weyl semimetals. Hence it protects surface state Fermi arcs which form an anomalous band structure consisting of disconnected curves that connect the projections of Weyl nodes with opposite topological charges on the boundary of a bulk sample. In contrast to the Dirac fermion states realized in graphene (*3*), topological insulators (*4-6*) and Dirac semimetals (*6, 12*), the Weyl semimetal state does not require any symmetry for its protection except the translation symmetry of the crystal lattice itself. The unique robustness of the Weyl fermions suggests that they could be exploited to carry electric currents more efficiently than normal electrons, thereby possibly pushing electronics to new heights. The first-ever realization of Weyl fermions in all physics, the new topological classification, the exotic Fermi arc surface states, the predicted quantum anomalies, and the prospects in device and applications suggest Weyl fermions and Weyl semimetals as the next era (after Dirac) in condensed matter physics (*8-17*).

Despite interest, Weyl semimetals are extremely rare in nature and, for many years, there was no successful realization of a Weyl semimetal in a real material. Recently, it was proposed that a family of isostructural compounds, TaAs, NbAs, TaP and NbP, are Weyl semimetals (*18, 19*). Shortly after the theoretical prediction, the first Weyl semimetal was experimentally discovered in TaAs (*20*). Around the same time, Weyl points were also observed in a non-fermionic system, a photonic crystal (*21*). The electronic Weyl system (TaAs) was experimentally observed via photoemission (*20, 22*). Later, other independent later experiments have confirmed the Weyl state in TaAs and shown the Weyl state in the other member, NbAs (*23-25*). It is important to note that both TaAs and NbAs contains arsenic. Here, we report the experimental observation of a Weyl semimetal without toxic elements in TaP.

We use soft X-ray angle-resolved photoemission spectroscopy (SX-ARPES) (*26, 27*) and ultraviolet ARPES to study the bulk and surface band structure, respectively, of TaP. In the bulk, we observe point degeneracies at arbitrary points in the Brillouin zone, with bands which disperse linearly in all directions in momentum space away from the degeneracy, which demonstrates Weyl points and Weyl cones in the bulk of TaP. We observe Fermi arc surface states on the surface of TaP. Our results are qualitatively consistent with first principles calculations, providing further support to our experimental data and demonstrating unambiguously that TaP is a Weyl semimetal. Moreover, we present an experimentally practical approach to demonstrating the bulk-boundary correspondence of a Weyl semimetal by resolving the net number of chiral edge modes on a closed path that encloses the Weyl node. Our results not only identify the Weyl semimetal state in TaP but also demonstrate a systematic experimental methodology that can be more generally applied to discover other Weyl semimetals and to measure their topological invariants.

**RESULTS**

Tantalum phosphide, TaP, crystallizes in a body-centered tetragonal Bravais lattice, space group $I4_1md$ (109), point group $C_{4v}$. An X-ray diffraction (XRD) measurement showed that the lattice constants of our TaP samples are $a = 3.32 \ \overset{\circ}{A}$ and $c = 11.34 \ \overset{\circ}{A}$, consistent with earlier crystallographic studies (*29*). The basis consists of two Ta atoms and two P atoms, as shown in Fig. 1A. In this crystal structure each layer is shifted relative to the layer below by half a lattice constant in either the *x* or *y* direction. This shift gives the lattice a screw pattern along the *z* direction, which leads to a non-symmorphic $C_4$ rotation symmetry that includes a translation along the $z$-direction by $c/4$. Crucially, we note that TaP lacks space inversion symmetry. The breaking of space inversion symmetry is a crucial condition for the realization of a Weyl semimetal because in the absence of inversion symmetry all bands are generically singly-degenerate. In Fig. 1D we show the bulk Brillouin zone and (001) surface Brillouin zone of TaP. In Figs. 1B and C we show the calculated band structure along some high symmetry lines. It can be seen that the conduction and valence bands cross each other near the $\Sigma$, $\Sigma'$, and $N$ points giving rise to a semimetal groundstate. In the absence of SOC it can be seen that the overlap of the conduction and valence bands yields four ring-like crossings, nodal lines, in the $k_x = 0$ and $k_y = 0$ mirror planes, as shown in Fig. 1E. When SOC is taken into account, the nodal lines dissolve into point band crossings, the Weyl nodes, which are shifted away from the $k_x = 0$ and $k_y = 0$ mirror planes. We show the positions of the Weyl points from calculation in Fig. 1E. The black and white circles represent Weyl points of opposite chiralities, the blue planes are mirror planes and the red plane is the $k_z = 2\pi/c$ plane. Our band structure calculation shows that there are 24 Weyl nodes, which can be categorized into two groups, W1 and W2. W1 represents the 8 Weyl nodes that are located on the $k_z = 2\pi/c$ plane, and W2 represents the remaining 16 Weyl nodes that are away from the $k_z = 2\pi/c$ plane. In Fig. 1F, we show the dispersion at the $k_x = 0$ mirror plane, $E(k_x = 0, k_y, k_z)$ around the nodal line (yellow line) when SOC is ignored. We note that the nodal line disperses in energy (Fig. 1F). Because the Weyl nodes emerge at $k$ points near the nodal line upon the inclusion of SOC, the dispersion of the nodal line leads to an energy offset between the W1 and W2 Weyl nodes. All W1 Weyl nodes project as single Weyl nodes in the close vicinity of the surface BZ edge $\overline{X}$ and $\overline{Y}$ points. Contrary to W1, pairs of W2 Weyl nodes with the same chiral charge project onto the same point of the (001) surface BZ. Moreover, the eight projected W2 Weyl nodes have projected chiral charge of ±2. The locations of the Weyl nodes and their projected chiral charge are schematically shown in Fig. 1H for the (001) surface for TaP (TaAs) and NbP (NbAs) in the left and right panel, respectively. An important difference between the left and right panel of Fig. 1H is their difference in the W1 chiralities. Interestingly, the projected chiral charge for the W1 Weyl nodes on the (001) surface is opposite for TaP (TaAs) and NbP (NbAs). This effect is not due to the $k$-space exchange of the W1 Weyl nodes chiral charge in the bulk. Instead, it arises from how the W1 is projected onto the (001) surface. As shown in Fig. 1G, it depends on whether the W1 is to the left or to the right of the $N$ point. In other words, there is no topological phase transition in the bulk by going from TaP to NbP.

However, on the surface, the projected W1 Weyl nodes will move around so it would be interesting to study how the Fermi arcs associated with the W1 Weyl node evolve in this process if they can be resolved experimentally.

We now present ARPES data to study the electronic structure of TaP. Specifically, we use SX-APRES ($hv = 350 - 1,000$ eV), which is reasonably bulk sensitive (*26, 27*) and therefore selective in $k_z$ (*28*), to measure the bulk band structure, and we use vacuum ultraviolet ARPES ($hv = 35 - 90$ eV), which is surface sensitive, to measure the surface band structure. We begin our discussion by presenting the measurements of the bulk Weyl cones using SX-ARPES. In Fig. 2A we show the SX-ARPES measured $k_z - k_x$ Fermi surface map over multiple BZs along the $k_z$ direction for a TaP crystal from Batch I. The observed $k_z$ dispersion in Fig. 2A, demonstrates that SX-ARPES is predominantly bulk sensitive. Figure 2B shows the SX-ARPES measured $k_x - k_y$ Fermi surface at the $k_z$ value corresponding to the W2 Weyl nodes by using an incident photon energy of 650 eV. In Fig. 2C, we show a $k_x - k_y$ calculation of the Fermi surface performed at the $k_z$ value of the W2 Weyl nodes. The chemical potential of the sample is estimated to be about 20 meV below the W2 Weyl node (further discussed below). The calculated Fermi surface is in qualitative agreement with the ARPES data. It can be seen that the Fermi pockets that arise from the two nearby W2 Weyl nodes are merged into one because the chemical potential is below the W2 Weyl nodes. Figures 2E,F show the energy dispersion near the W1 and W2 Weyl nodes along an in-plane momentum space cut direction that goes across a pair of W1 (W2) Weyl cones. The spectrum clearly shows a linear dispersion in the in-plane directions. Because the cuts to cross a pair of nearby cones are chosen, one would expect to see two Weyl nodes along each cut. However, for W1 (Fig. 2E), we cannot resolve the two nodes because they are too close to each other in $k$-space. For W2 (Fig. 2F), we also cannot resolve the two nodes possibly because the chemical potential is below the Weyl nodes. For the W1 Weyl cone observed in Fig. 2F, our data shows the upper Weyl cone. The W1 Weyl node is found to be about 40 meV below the Fermi level according to our data. Note that in our calculation, we find that there is a 64 meV offset in energy for the W1 and W2 Weyl nodes, shown in Fig. 2D. From these two facts, we know that the chemical potential is about 24 meV below the W2 Weyl nodes. Furthermore, because a Weyl cone disperses linearly in all three directions, it crucial to study the dispersion of the W2 Weyl bands along the out-of-plane $k_z$ direction. By fixing $k_x$ and $k_y$ at the location corresponding to the W2 Weyl node we are able to study the energy dispersion as a function of $k_z$. As shown in Fig. 2G, the Weyl cone is observed to have a strong $k_z$ dispersing character, which demonstrates its 3D nature. Around $k_z = \pm 1.3\pi/c$, the observed dispersion in the data is consistent with the linear dispersion along the $k_z$ direction, shown in Fig. 2H. In Fig. 2I, we report SX-ARPES measurements using a TaP sample from a different batch, Batch II, which is relatively more *n*-type than Batch I. Following the measurement procedure previously mentioned, the $E - k_x$ dispersion is also shown to be linear. Fig. 2I shows that the two W2 Weyl nodes form two linearly dispersing bands. We have shown that the bulk bands of TaP form discrete points at the Fermi level in the bulk BZ, that the bands disperse linearly along both in-plane and out-of-plane directions away from the band touching points, and that the band touchings are not located at any high-symmetry points or along any rotational axis. Our ARPES data are consistent with the bulk band calculations. All these together establish the Weyl cones and Weyl nodes in TaP.

Next, we examine the surface states by using low-energy ARPES. In Fig. 3A we show high-resolution ARPES Fermi surface map of the (001) surface of TaP at a photon energy of $h\nu = 50$ eV. At the Fermi level, our data shows a bowtie shaped feature centered at $\overline{X}$, a long elliptical shaped feature centered at $\overline{Y}$ and a tad-pole like feature that extends along each $\overline{\Gamma} - \overline{X}$ or $\overline{\Gamma} - \overline{Y}$ line. Figure 3B shows the theoretically calculated surface Fermi surface of the P termination. A concurrent photoemission study where the authors claimed that they observed surface states from the Ta termination (*30*). We observe a qualitative agreement between the data and the calculation. Because the pairs of W1 Weyl nodes are very close to each other in $k$-space, and because the bowtie and long-elliptical features show broader spectra than the tadpole features, resolving the Fermi arcs associated with the W1 Weyl nodes is beyond our experimental resolution. We note that this was also the case in both TaAs and NbAs (*20, 22, 24, 25*). We focus on the surface states near the W2 Weyl nodes because the k separation of the W2 is much larger and well within our experimental resolution. Figure 3C shows a zoomed-in high-resolution Fermi surface of the tadpole along the $\overline{\Gamma} - \overline{Y}$ line. From Fig. 3C, we observe that the tadpole is comprised of the following distinct features, going from $\overline{\Gamma} - \overline{Y}$: Two bright and one dark moon-shaped features, one closed inner contour, and a long tail-like feature. Figure 3D shows the corresponding zoomed-in Fermi surface in calculation, where a qualitative agreement can be seen. Specifically, in the calculation, going from $\overline{\Gamma} - \overline{Y}$, we also observe two bright moon-shaped features, a dark moon-shaped feature, one closed inner contour, and a long tail-like feature. The difference is the location of the closed inner contour. We also note that in the calculation there are projected bulk bands at the Fermi level in the $k$-space region of the inner contour, which makes the inner Fermi arc's spectral weight very weak when it approaches the $\overline{\Gamma}-\overline{Y}$ line in the calculation. We identify the arcs in calculation by further zooming near the neck region of the tadpole, shown in the inset of Fig. 3D. It can be seen that the second bright moon and the dark moon features are the arcs because they terminate onto the Weyl nodes. Other features are trivial surface states because they avoid the Weyl nodes, although the avoid crossing is quite small. Resolving the arcs in experiments by the same method is beyond our resolution because the Weyl nodes have very low spectral weight at low photon energies and because the surface states are very close to each other. Below in Fig. 4, we will experimentally demonstrate that among all of the surface states which we observe near the W2 Weyl nodes there are exactly two which must be Fermi arcs. Here, in Fig. 3E, we show a schematic configuration of the surface states that correspond to our data in Fig. 3C. The assignment of Fermi arcs and trivial surface states is made by comparison between our ARPES data (Fig. 3C) and calculation (Fig. 3D). We study the energy dispersion of the tadpole feature on the $\overline{\Gamma}-\overline{Y}$ line. As shown in Fig. 3F, our data show that there are six surface states associated with the tadpole feature crossing the Fermi level. We identify an important origin for the observed trivial surface states. Figures 4B,C show the calculated surface state dispersion along the $\overline{\Gamma}-\overline{Y}$ line. The calculation used the Green's function technique to obtain the spectral weight from the top region of a semi-infinite TaP system. Figure 4B shows the surface spectral weight only from the top unit cell whereas Figure 4C considers the top two unit cells. It can be seen that the inclusion of the second layer leads to additional two surface states that cross the Fermi level, as pointed out by the white arrows in Fig. 4C. Therefore, we believe that one of the important origins for the trivial surface states can be the contribution from deeper layers.

**DISCUSSIONS**

We discuss the methods that have been used to demonstrate the existence of Fermi arcs. First, according to the literal definition of a Fermi arc, one needs to observe a disjoint segment (a non-closed curve) that terminates at two Weyl nodes. This can be used in calculations where resolution is sufficient and indeed this is the case for our calculation presented in the inset of Fig. 3D. However, in experiments, this is extremely difficult for the TaAs class of compounds because the bulk bands have very low spectral weight at low photon energies and because there are many surface states (both topological and trivial) in the vicinity of each Weyl node. Second, it was proposed in Refs. (*22, 25*) that one can demonstrate the existence of Fermi arcs in the TaAs class of compounds by observing an odd number of surface state Fermi crossings along the the $k$-space path $\overline{\Gamma} - \overline{X}(\overline{Y}) - \overline{M} - \overline{\Gamma}$. It was argued (*22*) that that this is desirable because one can simply count the number of crossings along these high symmetry lines without worrying about the details of the surface states in the close vicinity of the Weyl nodes. These proposed loops are shown by the green and magenta triangles in Fig. 4D. Here we show that although this method is theoretically correct, it can be quite impractical experimentally. The red and blue lines in Fig. 4D show a surface state configuration that is allowed by the projected Weyl nodes and their chiral charges through the bulk-boundary correspondence. Indeed, theoretically, going around the green or magenta triangle should give an odd number of crossings. However, one should note that the pairs of W1 Weyl nodes are very close to each other in $k$-space. Hence, in Fig. 4D, the Fermi arcs (the red lines) near the $\overline{X}$ point are very short. Similarly, around the $\overline{Y}$ point in Fig. 4D, the openings between the two Fermi arcs are very narrow. Therefore, obtaining an odd number of Fermi crossings along the green triangle means that one must experimentally resolve the band crossing that arises from the very short arc near the $\overline{X}$ point. Similarly, obtaining an odd number of Fermi crossings along the magenta triangle means that one must experimentally resolve the opening between the two Fermi arcs (two red lines) near the $\overline{Y}$ point. In other words, observing an odd number of Fermi crossings along these paths is equivalent to directly resolving the Fermi arcs associated with the W1 Weyl nodes near the $\overline{X}$ or $\overline{Y}$ point. Because the W1 Weyl nodes are very close to each other in $k$-space and because the surface states near the $\overline{X}$ and $\overline{Y}$ points have a broad spectrum, resolving their arc character seems to be quite challenging in experiments.

Here we propose a method to demonstrate the existence of Fermi arcs that is not only theoretically rigorous but also experimentally feasible. We propose to count the net number of chiral edge modes on a closed path that encloses the Weyl node. Specific to the case of TaP (also TaAs and NbAs), we propose the rectangular loop that encloses only one projected W2 Weyl node, as shown in Fig. 4E. In the bulk BZ, the rectangular loop corresponds to a rectangular pipe that encloses two W2 Weyl nodes of chiral charge +1. Therefore, according the topological band theory, the Chern number on this manifold has to be two, which implies that there must be two net chiral modes along this rectangular loop. Figures 4F-I shows the ARPES spectrum following the rectangular path in a counter-clockwise fashion, from I to IV. We see that in total there are four left moving surface modes and two right moving surface modes. Therefore, our ARPES data in Figs 4F-I shows that there are two net chiral edge modes. This means that among the six observed Fermi crossings, two of them have to be Fermi arcs and the remaining four form closed contours. Theoretically, Fermi arcs are expected to terminate onto Weyl nodes while trivial surface states should avoid them. However, the avoided crossing of trivial surface states with Weyl nodes can be very small in real samples, which is indeed the case for TaP. As seen in Fig. 4E, these surface states are very close to each other at the location of the W2 Weyl nodes (the black and white circles are placed based on the theoretically calculated locations of the W2 Weyl nodes). Thus experimentally resolving which two out of the six Fermi crossings are the Fermi arcs is not experimentally practical, but at least we unambiguously demonstrate the presence of two Fermi arcs connecting to a projected W2 Weyl node. We note that in order to establish the Weyl semimetal state, demonstrating the existence of one set of Fermi arcs is sufficient. The method described above can also be applied to show Fermi arcs in the theoretical calculations. For example, the white dotted rectangular loop in the inset of Fig. 3D encloses one projected W2 Weyl node with the projected chiral charge of 2. Along this loop, there are four Fermi crossings. Three disperse to one direction while the other one disperses to the opposite direction, which also leads to same result, two net chiral edge modes. Therefore, both the experimental data and the theoretical calculations give consistent results, which are further consistent with the topological band theory. If one chooses the loop used in the experimental data, the counting is more complicated because the inner arc becomes very weak at $k$ points along the loop due to overlap with the bulk projection. We further note that the number of surface state crossings around a projected W2 Weyl node should always equal 2 + an even number (the inset of Fig. 4E). Here, the first term is required by the projected chiral charge of 2 and the second term represents the fact that additional trivial contours will always contribute an even number of additional crossings. Therefore, around a projected W2 Weyl node there should always be an even number of surface states, and an odd number of surface states would violate topological band theory. We note that the proposed rectangular loop is experimentally feasible because it only encloses the W2 and not the W1 Weyl node. Therefore, we do not need to resolve the surface states near the W1 nodes, which is experimentally difficult due to their close proximity. Finally, we discuss the difference between TaP and another prototypical Weyl semimetal TaAs (*20*). A prominent distinction is that we found that TaP has a number of additional trivial surface states, closed contours, besides the Fermi arcs, which were not the seen in TaAs (*20*). Since in both compounds the termination consists of the pnictide (As or P) atoms, the additional surface states suggest that the P surface environment is different from that of As. Therefore, it would be interesting to study the evolution of surface state in $TaP_{1-x}As_x$. It would be also interesting to change the surface environment in TaP via surface deposition or adsorption to study the interplay between the Fermi arc and the trivial surface states. In certain conditions, a topological Fermi arc and a trivial surface state can switch their roles, while the constraint placed by topology that the net number of topological Fermi arcs cannot change must be met.

We summarize our experimentally-practical method for demonstrating Fermi arcs and measuring the topological invariants of Weyl semimetals, as proposed in this study. We choose a closed path in $k$-space and subtract the number of right-moving modes from the number of left moving modes. The result equals the total chiral charge of the Weyl nodes enclosed by the path. In real experiments, one should consider the sharpness of the features and the distance between the Weyl nodes in different regions of the Brillouin zone, and choose the $k$-path based on these factors. In summary, we have presented systematic ARPES data that reveals the bulk and surface electronic structure of the Weyl semimetal candidate TaP. Our results have identified TaP as the first Weyl semimetal that contains no toxic elements. The non-toxic property is crucial for future research studies and potential applications. We have also demonstrated, for the first time, a systematic experimental methodology that can be more generally applied to discover other Weyl semimetals and to measure their topological invariants.

**MATERIALS AND METHODS**

Sample growth and ARPES measurement techniques: High quality single crystals of TaAs were grown by the standard chemical vapor transport method reported in Ref. (*20*). High resolution vacuum ultraviolet angle-resolved photoemission spectroscopy (ARPES) measurements were performed at Beamlines 4.0.3, 10.0.1 and 12.0.1 of the Advanced Light Source (ALS) at the Lawrence Berkeley National Laboratory (LBNL) in Berkeley, California, USA, Beamline 5-4 of the Stanford Synchrotron Radiation Light source (SSRL) at the Stanford Linear Accelerator Center (SLAC) in Palo Alto, California, USA and Beamline I05 of the Diamond Light Source (DLS) in Didcot, UK, with the photon energy ranging from 15 eV to 100 eV. The energy and momentum resolution of the vacuum ultraviolet ARPES instruments was better than 30 meV and 1% of the surface Brillouin zone (BZ). The SX-ARPES measurements were performed at the Adress Beamline at the Swiss Light Source in the Paul Scherrer Institut (PSI) in Villigen,

Switzerland. The experimental geometry of SX-ARPES has been described in (*26-28*). The sample was cooled down to 12K to quench the electron-phonon interaction effects reducing the *k*-resolved spectral fraction. Our SX photon energy ranged from 300 to 1000 eV. The combined (beamline and analyzer) experimental energy resolution of the SX-ARPES measurements varied between 40 and 80 meV. The angular resolution of the SX-ARPES analyzer was 0.07°. Samples were cleaved *in situ* under a vacuum condition better than $5\times10^{-11}$ torr at all beamlines.

Computational calculation methods: First-principles calculations were performed by the OPENMX code based on norm-conserving pseudopotentials generated with multi-reference energies and optimized pseudoatomic basis functions within the framework of the generalized gradient approximation (GGA) of density functional theory (DFT) (*31*). Spin-orbit coupling was incorporated through *j*-dependent pseudo-potentials. For each Ta atom, three, two, two, and one optimized radial functions were allocated for the *s*, *p*, *d*, and *f* orbitals (*s3p2d2f1*), respectively, with a cutoff radius of 7 Bohr. For each P atom, *s3p3d3f2* was adopted with a cutoff radius of 9 Bohr. A regular mesh of 1000 Ry in real space was used for the numerical integrations and for the solution of the Poisson equation. A *k* point mesh of $17\times17\times5$ for the conventional unit cell was used and experimental lattice parameters were adopted in the calculations. Symmetry-respecting Wannier functions for the P *p* and Ta *d* orbitals were constructed without performing the procedure for maximizing localization and a real-space tight-binding Hamiltonian was obtained (*32*). This Wannier function based tight-binding model was used to obtain the surface states by the recursive Green's function method.

**Acknowledgments:** Work at Princeton University and Princeton-led synchrotron-based ARPES measurements were supported by the Gordon and Betty Moore Foundations EPiQS Initiative through Grant GBMF4547 (Hasan). Single crystal growth was supported by National Basic Research Program of China (Grant Nos. 2013CB921901 and 2014CB239302) and characterization by U.S. DOE DE-FG-02-05ER46200. First-principles band structure calculations at National University of Singapore were supported by the National Research Foundation, Prime Minister's Office, Singapore under its NRF fellowship (NRF Award No. NRF-NRFF2013-03). T.-R.C. and H.-T.J. were supported by the National Science Council, Taiwan. H.-T.J. also thanks National Center for High-Performance Computing (NCHC), Computer and Information Network Center National Taiwan University (CINC-NTU), and National Center for Theoretical Sciences (NCTS), Taiwan, for technical support. F.C.C acknowledges the support provided by MOST-Taiwan under project number 102-2119-M-002-004. We gratefully acknowledge J. D. Denlinger, S. K. Mo, A. V. Fedorov, M. Hashimoto, M. Hoesch, T. Kim, and V. N. Strocov for their beamline assistance at the Advanced Light Source, the Stanford Synchrotron Radiation Lightsource, the Diamond Light Source, and the Swiss Light Source. We also thank D. Huse, I. Klebanov, A. Polyakov, P. Steinhardt, H. Verlinde, and A. Vishwanath for discussions. R.S. and H.L. acknowledge visiting scientist support from Princeton University. Requests for data or other materials should be addressed to M.Z.H. (mzhasan@princeton.edu).

**Author contributions:** S.-Y.X., I.B., N.A., D.S.S, and G.B. conducted the ARPES experiments with assistance from N.A. H.Z., P.V.S. M.L.P. V.N.S., and M.Z.H.; C.G., C.Z., Z.Y., Y.F., H.L., R.S., F.C.C., and S.J. grew the single-crystal samples; G.C., T.-R.C., C.-C.L., S.-M.H., C.-H.H, H.-T.J., A.B., and H.L. performed first-principles band structure calculations; T.N. did theoretical analyses; M.Z.H. was responsible for the overall direction, planning and integration among different research units.

**Competing interests**

The authors declare that they have no competing interests.

**Figures**

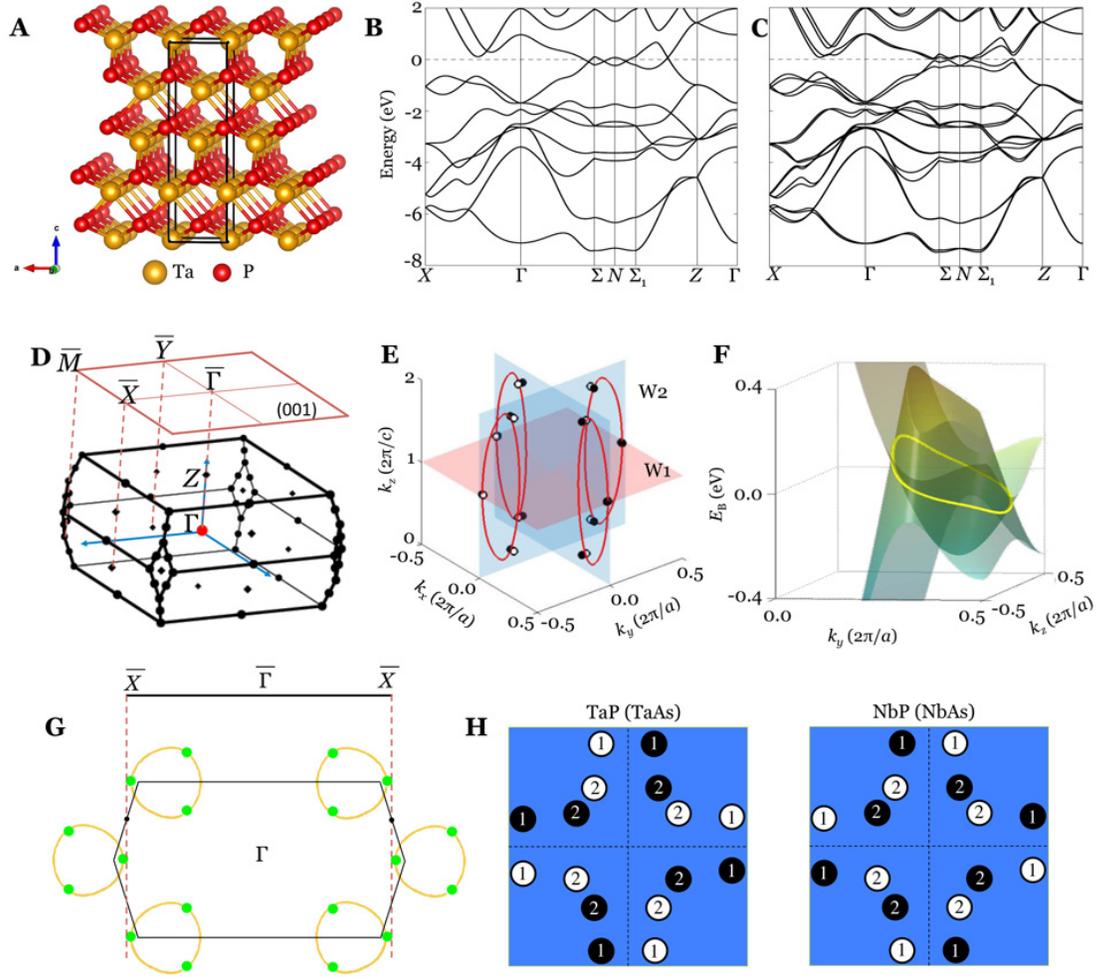

**Fig. 1. Electronic band structure of the Weyl semimetal TaP.** (**A**) Body-centered tetragonal structure of TaP, shown as stacks of Ta and P layers. (**B**) First-principles band structure calculation of the bulk TaP without spin-orbit coupling. (**C**) Same as panel (B) but with spin-orbit coupling. Schematic of the distribution of Weyl nodes in the three-dimensional Brillouin zone (BZ) of TaP. (**D**) The bulk Brillouin zone and (001) surface Brillouin zone of TaP, with certain high-symmetry points labeled. (**E**) Calculation of the positions of the Weyl nodes, with opposite chiralities indicated by the white and black circles. The mirror planes are blue and the $k_z = 2\pi/c$ plane is red. (F) The band structure on the $k_x = 0$ mirror plane, $E(k_x = 0, k_y, k_z)$, in the vicinity of the ring-like crossing, in yellow, which we find in calculation when SOC is ignored. (**G**) Calculation showing the ring-like crossings in the $k_x = 0$ plane, with the Weyl nodes indicated by green dots. The distribution of chiralities of the projected W1 nodes in the first surface

Brillouin zone depends on whether the ring-like crossing is large enough that the W1 nodes spill over the edge of the first surface Brillouin zone, marked by the dotted line through the bulk N point. (**H**) To understand the distribution of chiralities more clearly, we show cartoons (not to scale) of the projected Weyl nodes and their chiral charges on the (001) Fermi surface of TaP (TaAs) and NbP (NbAs), respectively. The projected Weyl nodes are denoted by black and white circles, their color indicates their opposite chiralities and the number in the circle indicates the projected chiral charge. We find that the chiralities of the W1 points are swapped in TaP (TaAs) with respect to NbP (NbAs).

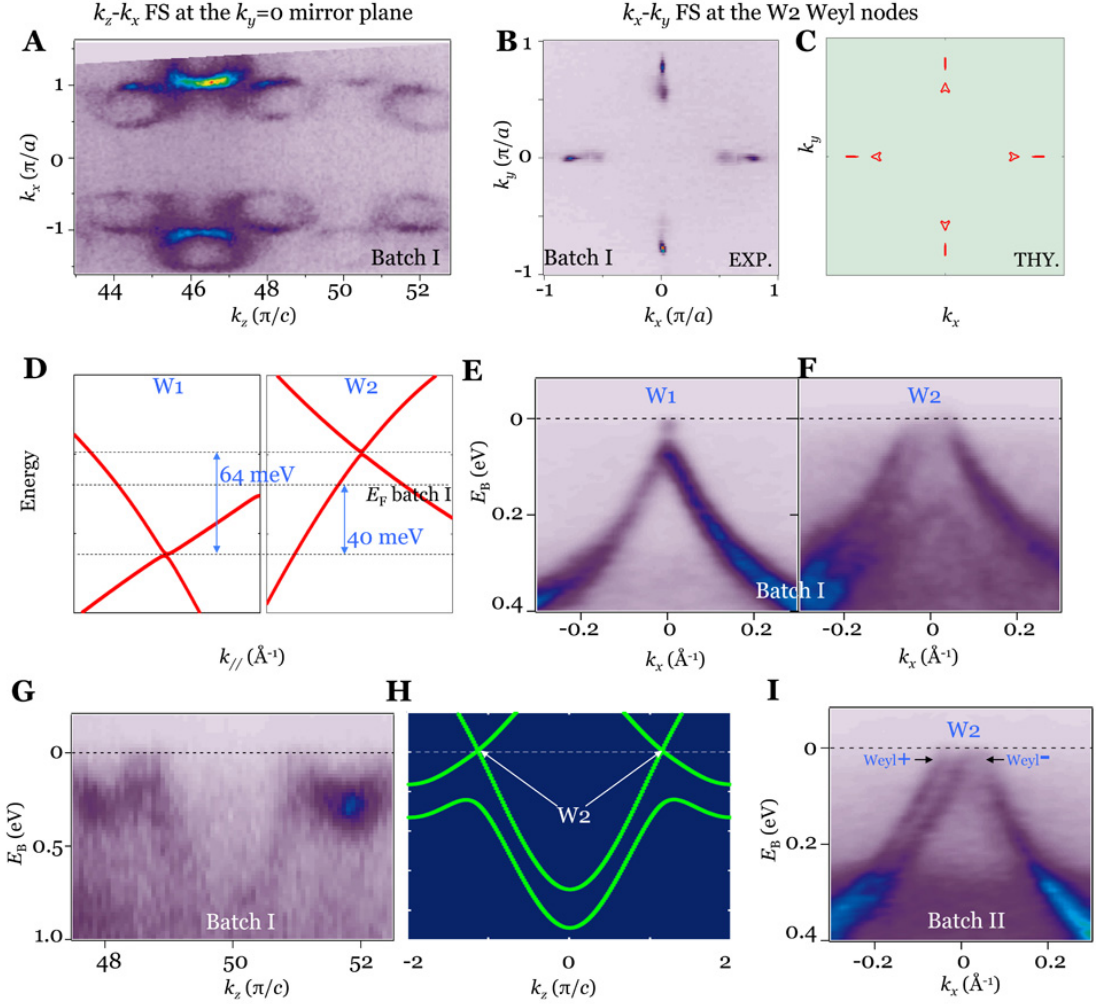

**Fig. 2. Bulk Weyl fermion cones in TaP.** (**A**) SX-ARPES Fermi surface map at $k_y = 0$ in the $k_z - k_x$ plane. (**B**) SX APRES Fermi surface map at $k_z = W_2$ in the $k_x - k_y$ plane, showing the W2 Weyl nodes. (**C**) Theoretical calculation of the same slice of the Brillouin zone at the same energy shows complete agreement with the soft X-ray ARPES measurement. (**D**) First-principles calculations of the energy-momentum dispersion of the two sets of Weyl nodes, W1 and W2, in TaP. The two nodes are offset in energy by 64 meV. (**E,F**) Energy dispersion maps, $E - k_{//}$ for W1 and W2, respectively. (**G,H**) ARPES measured and theoretically calculated out-of-plane dispersion, $E - k_z$ for W2, which shows the linear dispersion of the Weyl cone along the out-of-plane direction. All data in panels (A-G) are obtained from Batch I. (**I**) Energy dispersion of the W2 Weyl cones from a sample in Batch II. We see the $+/-$ Weyl nodes more clearly for Batch II TaP, as labeled in panel (I).

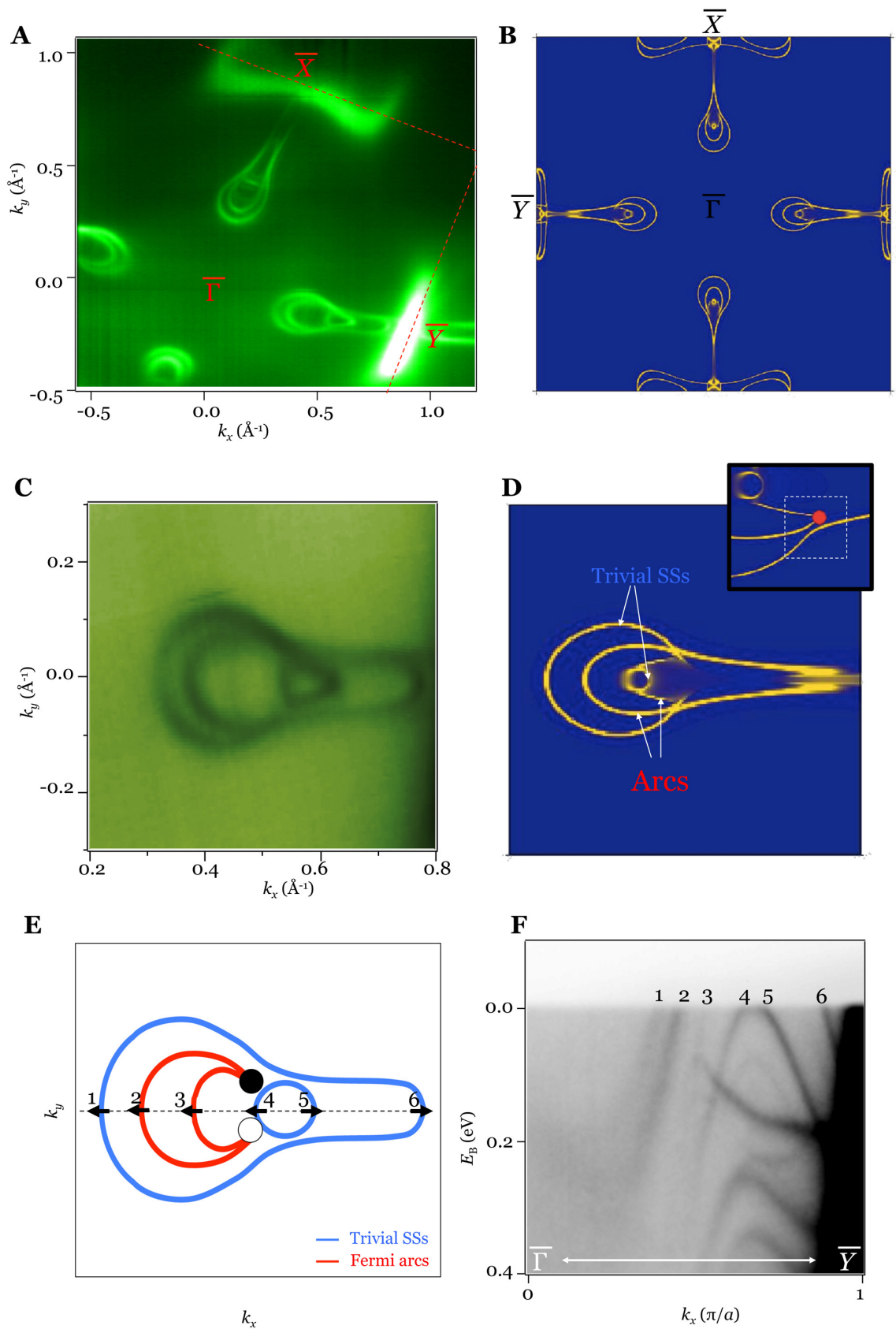

**Fig. 3. Fermi Arc surface states in TaP.** (**A,B**) APRES measured Fermi surface and first-principles band structure calculation of the (001) Fermi surface of TaP, respectively. (**C**) High-resolution ARPES Fermi surface map in the vicinity of the $\overline{\Gamma}-\overline{Y}$ high symmetry line. (**D**) Theoretical calculated surface Fermi surface along $\overline{\Gamma}-\overline{Y}$. The calculation used the Green's function technique to obtain the spectral weight from the top two unit cells of a semi-infinite TaP system. (**E**) Schematic showing the Fermi arcs and the trivial surface states that correspond to our data in panel (C). This configuration is obtained by analyzing our ARPES data and comparing it with calculations (see main text). (**F**) ARPES dispersion along the $\overline{\Gamma}-\overline{Y}$ high symmetry line. The six Fermi crossings are numbered 1-6. We see 4 states (1-4) with one sign of Fermi velocity and the other two (5 and 6) with the opposite sign. This is consistent with projected chiral charge ±2 for the W2. These six states are also labeled in panel (E), where the arrows represent their corresponding Fermi velocity direction.

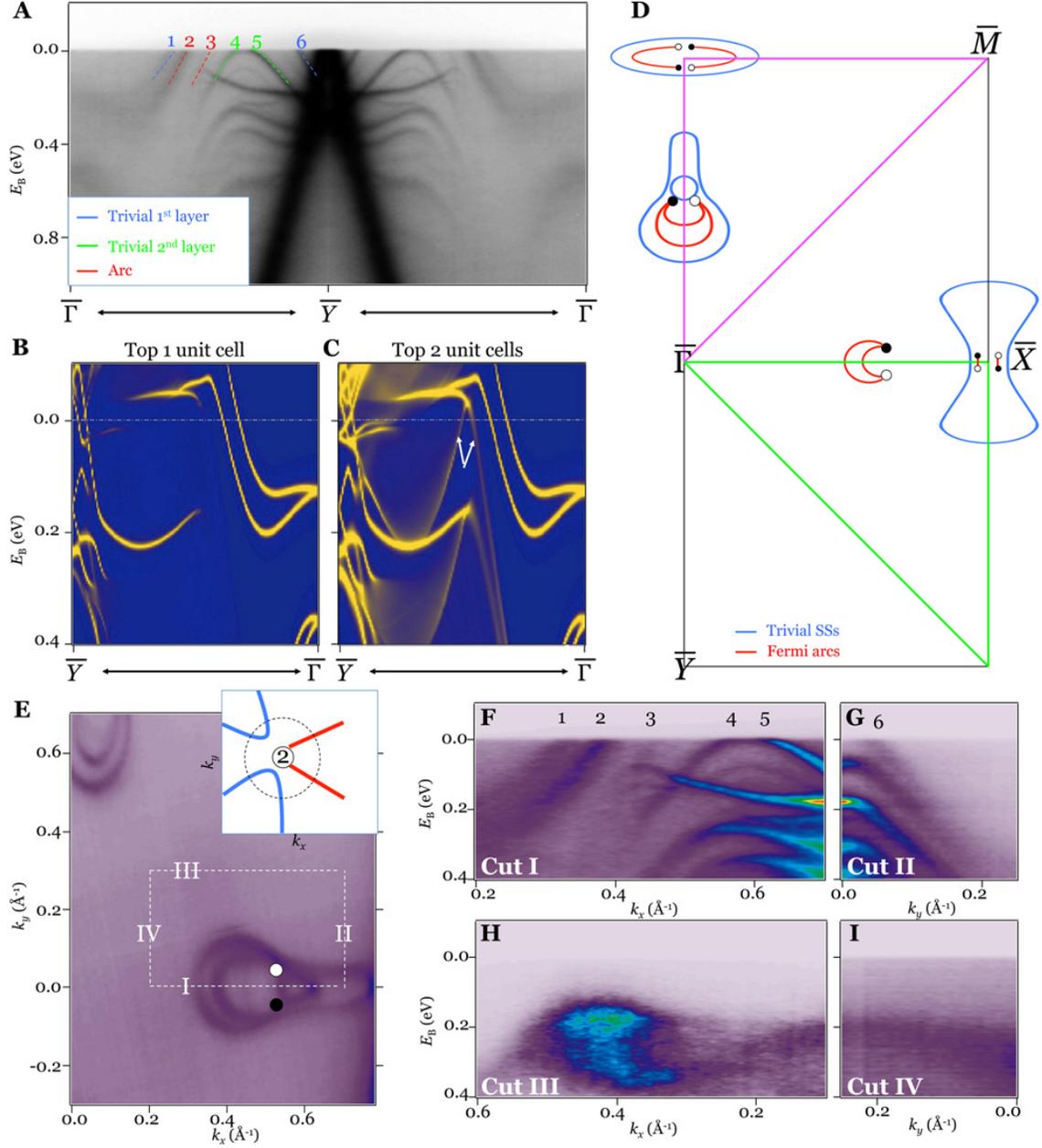

**Fig. 4. Bulk boundary correspondence and topological nontrivial nature of TaP.** (**A**) ARPES energy-momentum dispersion map along the $\overline{\Gamma}-\overline{Y}-\overline{\Gamma}$ high-symmetry line, which shows six band crossings the Fermi level. (**B**) Calculated energy dispersion along the $\overline{Y}-\overline{\Gamma}$ line. The calculation uses the Green's function technique to obtain the spectral weight from the top one unit cell of a semi-infinite TaP system. (**C**) Similar to panel (B) but for the top two unit cells. (**D**) Schematic examples of closed paths (green and magenta triangles) that enclose both W1 and W2 Weyl nodes. The bigger black and white circles represent the projected W2 nodes whereas the smaller ones correspond to the W1 nodes. The red lines denote the Fermi arcs and the blue lines show the trivial surface states. The specific configuration of the surface states in this cartoon is not intended to reflect the Fermi arc connectivity in TaP. It simply provides an example of one configuration of the surface states that is allowed by the projected Weyl nodes and their chiral charges by the bulk-boundary correspondence. (**E**) Zoomed in ARPES spectra near the W2 nodes along $\overline{\Gamma}-\overline{Y}$. The white-dashed line marks a rectangular loop that encloses a projected W2 Weyl node. The circles are added by hand based on the calculated locations of the projected W2 nodes. (**F-I**) ARPES dispersion data as one travels around the k space rectangular loop shown in panel (E) in a counter-clockwise way from I to IV. The closed path has edge modes of net chirality two, showing that the path must enclose a projected chiral charge of two using no first-principles calculations, assumptions about the crystal lattice or band structure, or ARPES spectra of the bulk bands.